\documentclass{emulateapj}
\usepackage{graphicx}
\usepackage{natbib}
\pdfoutput=1
\usepackage{amsmath,amssymb}
\newcommand{\her}{$\it{Herschel~}$}

\newcommand{\beq} {\begin{equation}} \newcommand{\eeq}
{\end{equation}} \newcommand{\beqa} {\begin{eqnarray}}
\newcommand{\eeqa} {\end{eqnarray}}

\newcommand{\obs} {{\rm obs}}

 \newcommand{\krho} {{k_\rho}}
 
\newcommand{\lt} {\tilde L}

\newcommand{\rest}{{\rm rest}}
\newcommand{\rch} {R_{\rm ch}}
\newcommand{\rct} {{\tilde R_c}}

\newcommand{\rmt} {{\tilde r_m}}

\newcommand{\tch} {T_{\rm ch}}

\begin{document}


\bibliographystyle{apj}

\title{Photometric Redshifts of Submillimeter Galaxies}

\author{Sukanya Chakrabarti\altaffilmark{1}, Benjamin Magnelli\altaffilmark{3}, Christopher F. McKee\altaffilmark{2}, Dieter Lutz\altaffilmark{3}, Stefano Berta\altaffilmark{3},Paola Popesso\altaffilmark{3} \& Francesca Pozzi\altaffilmark{4,5}}
\altaffiltext{1}{Physics Dept, 777 Glades Road, Florida Atlantic University, Boca Raton, FL 33431 USA; schakra1@fau.edu}
\altaffiltext{2} {Physics And Astronomy Depts., UC Berkeley, Berkeley, CA 94720}
\altaffiltext{3}{Max-Planck-Institut f\"ur Extraterrestrische Physik, Postfach
1312, Giessenbachstrasse 1, 85741 Garching, Germany} 
\altaffiltext{4}{Dipartimento di Astronomia, Universit\`a degli Studi di Bologna, Via Ranzani 1, I40127 Bologna, Italy}
\altaffiltext{5}{INAF Osservatorio Astronomico di Bologna, Via Ranzani 1, I40127 Bologna, Italy}


\begin{abstract}

We use the photometric redshift method of Chakrabarti \& McKee (2008) to infer photometric redshifts of submillimeter galaxies with far-IR (FIR) $\it{Herschel}$ data \footnote{\textit{Herschel} is an ESA space observatory with science instruments
provided by European-led Principal Investigator consortia and with important
participation from NASA} obtained as part of the PACS Evolutionary Probe (PEP) program.  For the sample with spectroscopic redshifts, we demonstrate the validity of this method over a large range of redshifts ($ 4 \ga z \ga 0.3$) and luminosities, finding an average accuracy in $(1+z_{\rm phot})/(1+z_{\rm spec})$ of 10 \%.   Thus, this method is more accurate than other FIR photometric redshift methods.  This method is different from typical FIR photometric methods in deriving redshifts from the light-to-gas mass ($L/M$) ratio of infrared-bright galaxies inferred from the FIR spectral energy distribution (SED), rather than dust temperatures.   Once the redshift is derived, we can determine physical properties of infrared bright galaxies, including the temperature variation within the dust envelope, luminosity, mass, and surface density.  We use data from the GOODS-S field to calculate the star formation rate density (SFRD) of sub-mm bright sources detected by AzTEC and PACS.  The AzTEC-PACS sources, which have a threshold $850~\micron$ flux $\ga 5~\rm mJy$, contribute 15\% of the SFRD from all ULIRGs ($L_{\rm IR} \ga 10^{12} L_{\odot}$), and 3 \% of the total SFRD at $z \sim 2$.
\end{abstract}


\section{Introduction}

Much of the energy emitted by the universe in its infancy was from dust-enshrouded luminous galaxies.   The $\it{Herschel}$ Space Telescope has opened a new window into this epoch, yielding unprecedented sensitivity in the far-infrared (Pilbratt et al. 2010), where dusty galaxies emit most of their radiation.  The FIR SED allows us to probe the physical conditions of dusty sources.  Chakrabarti \& McKee (2005; henceforth CM05) developed an analytic means of self-consistently solving the radiative transfer equation for spherically symmetric, centrally heated dusty sources.  They derived a simple and intuitive form for the emergent SED that can be applied to infer the physical parameters of dust envelopes from the observed FIR SED.   Chakrabarti \& McKee (2008; henceforth by CM08 we refer to the paper and by "CM" we refer to CM08's photometric redshift method) applied the method of CM05 to fit observed FIR SEDs of ULIRGs and SMGs and showed that accurate photometric redshifts could be inferred from derivation of the light-to-gas mass ($L/M$) ratio.  

CM08 demonstrated the accuracy of their method with a sample of SMGs with FIR data from Kovacs et al. (2006).  This was the only FIR sample of SMGs available at the time, which constituted a sub-sample of the bright SMGs studied by Chapman et al. (2005).  The new PACS (Poglitsch et al. 2010) observations of SMGs (Magnelli et al. 2012) provide an ideal sample for testing the CM method on a larger and more diverse sample.  SMGs are high-redshift galaxies ($z \ga 0.3$) classified on the basis of their submillimeter flux.  "Classical" SMGs have $F_{850~\micron} \ga 5~\rm mJy$, which corresponds to a luminosity of $\sim3\times10^{12}L_{\odot}$ at $z \sim 2$.  ULIRGs are generally taken to refer to galaxies emitting $\ga 10^{12}L_{\odot}$ in the infrared (Soifer et al. 1984).    CM08 had noted that local ULIRGs have higher $L/M$ values than their high-redshift cousins, a point that we discuss further in \S 3.  The galaxies in the Magnelli sample cover four blank fields (GOODS-N, GOODS-S, COSMOS, and Lockman Hole) as well as a number of lensing clusters.  They span a range in luminosities from $7 \times 10 ^{13} L_{\odot} \ga L \ga 5\times 10^{11}L_{\odot}$, and redshifts between $\sim 0.3 - 5$.   Many of these sources have spectroscopic redshifts.  Thus, this sample is diverse enough (Magnelli et al. 2011b) to yield a very robust measure of the accuracy of the CM photometric redshift method.  This is an important test, as photometric redshifts will be crucial for analyzing the vast bulk of the observations expected from \textit{Herschel}.  Having derived photometric redshifts, we then calculate the star formation rate density of sources detected by AzTEC and PACS as a function of redshift in the GOODS-S field, which is homogeneously covered in the sub-mm down to the classical SMG threshold ($F_{850~\micron} > 5~\rm mJy$).

On the basis of IRAS observations, Sanders et al. (1988) suggested that local ULIRGs may be produced by the merger of gas-rich spirals.  The development of hydrodynamical codes that adequately model the collision of gas-rich spirals and the subsequent starburst that results from a violent merger (Springel et al. 2005) made it possible to test this suggestion, which had been anticipated early on by Toomre \& Toomre (1972).  The infrared emission of ULIRGs and SMGs during their life cycles was subsequently calculated using three-dimensional self-consistent radiative transfer calculations through the time outputs of SPH simulations of merging spirals with central AGN (Chakrabarti et al. 2007; Chakrabarti et al. 2008; Chakrabarti \& Whitney 2009) .  These calculations reproduced empirically derived correlations, such as the correlation between the ratio of the $25~\micron$ to $60~\micron$ fluxes and energetically active AGN (de Grijp et al. 1984; Chakrabarti et al. 2007).  Chakrabarti et al. (2008) reproduced the clustering of sources in $\it{Spitzer}$ IRAC color-color plots (Lacy et al. 2004), and explained that it was due to the prevalence of the starburst phase in the time evolution of these sources.   SMGs formed during these simulations have diverse properties (Hayward et al. 2012) and constitute a heterogeneous group.  However, the simulations analyzed in these papers were not cosmological simulations, but rather simulated binary mergers of gas-rich systems with central black holes that yield high star formation and accretion rates onto the central AGN (Springel et al. 2005).  Thus, the number density of the SMG population as a function of redshift could not be derived.  (See however Hayward et al. 2011 for a phenomenological derivation of the number density of SMGs in this context) Recently, Dave et al. (2010) performed hydrodynamical cosmological simulations and identified SMGs as the most rapidly star forming systems that match the observationally determined stellar mass function of SMGs.  In these simulations, SMGs  sit at the centers of large potential wells, and accrete gas-rich satellites, but are not typically undergoing major mergers.  However, Magnelli et al. (2010; 2012) noted that their high star formation rates for the bright SMGs ($SFR \sim 1000 M_{\odot}~\rm yr^{-1}$) are difficult to reconcile with Dave et al.'s (2010) simulations that produce SMGs with star formation rates lower by $\sim 3$ than what is inferred observationally.   Thus, models of SMGs cannot yet fully account for their contribution to the cosmic stellar mass assembly.

Using our method to derive photometric redshifts from the FIR SED can potentially yield a robust measure of the number density of SMGs and their contribution to the star formation rate density of the universe (and thereby their contribution to the cosmic stellar mass assembly), without requiring model dependent assumptions.  Deriving fundamental parameters of the brightest infrared galaxies, such as their luminosity, star formation rate, and dust temperature, cannot be accomplished without first deriving their redshifts.  Unlike optical studies (Madau et al. 1998; Hopkins 2004), the derivation of star formation rates from the infrared luminosity is not affected by extinction, thereby yielding the most robust measure of the contribution of dusty galaxies to the star formation rate density of the early universe.

The paper is organized as follows.  In \S 2.1 and \S 2.2, we review very briefly the formalism of CM05 with attention to the intuitive form of the emergent SED, and the redshift inference method presented in CM08.  In \S 3, we present the observational sample.  In \S 4, we demonstrate the accuracy of the method with respect to the \her sample with spectroscopic redshifts.  In \S 4.1, we apply our method to sources detected by AzTEC and PACS in the GOODS-S field and present their star formation rate density as a function of redshift.  We discuss future work and present caveats in \S 5, and conclude in \S 6.

\vspace{0.1in}
\section{Photometric Redshifts:  Theory}

\subsection{SEDs of Dusty Sources}
In CM05, we formulated an analytic solution for the FIR SEDs of
spherically symmetric, homogeneous dusty sources with a central source
of radiation.  We considered envelopes that emit most of their
radiation at wavelengths longer than 30 $\mu\rm m$, and are sufficiently opaque
that emission from the dust destruction front does not
significantly influence the FIR SED.  We defined characteristic parameters, $\rch$ and $\tch$ that
are analogous to the Rosseland photospheric radius and temperature
respectively, such that 
\beq L\equiv 4\pi\lt \rch^{2}\sigma \tch^{4}\;,
\label{eq:L} \eeq 
where $\sigma=5.67\times 10^{-5} \rm
~erg~cm^{-2}~s^{-1}~K^{-4}$ is the Stefan-Boltzmann constant.  $\lt$
is a number of order unity that allows for better agreement
with the numerical solutions, and is relevant for very extended atmospheres,
which have $\rct \gg 1$, reflecting the effective increase in emitting
area.  We find that
\begin{equation} \tilde{L}=1.6\rct^{0.1} \;
\label{eq:ltil_val}
\end{equation} 
is accurate to within $\sim$ 10 \% for dust envelopes with density scaling as $\rho \propto r^{-\krho}$ with $1\la \krho\la 2$, and $\rct\equiv R_{c}/\rch$.
Here, $R_{c}$ is the outer radius of the dust envelope; our relations apply for $\rct > 1$.

We assume that the characteristic frequency close to which the dust envelope emits 
most of its radiation, $\nu_{\rm ch}$, is
within the frequency range where the opacity is approximately a power
law: 
\beq
\kappa_{\nu}=\kappa_{\nu_{0}}(\nu/\nu_{0})^{\beta}~~~~~(30~\mu{\rm m
\la \lambda \la 1~mm}),
\label{eq:kappa}
\eeq
with $\beta=2$ (Weingartner \& Draine 2001).  For $\nu_0=3$~THz, corresponding to $\lambda_0=100~\micron$ (fiducial values chosen for convenience),
we adopt an opacity per unit mass of gas of
\beq
\kappa_{100\,\micron}=0.54\delta ~~~\mbox{cm$^2$ g$^{-1}$}.
\eeq
For $\delta=1$, this is twice the value given by the
Weingartner \& Draine (2001) (henceforth WD01) model for dust in the diffuse interstellar medium
since  we assume that grains in star-forming regions have ice mantles that
double the FIR opacity.
The WD01 opacity is based on a gas-to-dust mass
ratio of $M/M_d=105.1$; since the ice mantles most likely have
a different opacity per unit mass than the WD01 grains, we
do not attempt to infer the dust mass in the sources.
Deviations from solar metallicity, or from the
assumed dust model, can be taken into account by choosing
a different value for $\delta$.

The SED variable $\rct$ measures the extent of the dust envelope
probed by the far-IR SED, which is reflected in the shape of the SED:  very opaque envelopes have $\rct\sim 1$ (close to a blackbody), whereas
less dense envelopes can have $\rct \gg 1$ (very broad SED).   We model the emergent spectrum by assuming that 
the emission in each
frequency channel comes from a shell of thickness $\Delta r_{m}(\nu)$
centered at a radius $r_{m}(\nu)$, with a source function
$(2h\nu_{ch}^{3}/c^{2})\exp\left[{-h\nu/kT(\rmt)}\right]$ located at
an optical depth $\tau_\nu(\tilde{r}_{m})$:
\begin{equation}
\begin{split}
 F_{\nu} \propto \left(\frac{2h\nu_{ch}^{3}}{c^{2}}\right)
\tilde{\kappa}_{\nu}\tilde{\nu}^{3}(k_{\rho}-1)\tilde{r}_{m}^{2-k_{\rho}}
\Delta\tilde{r}_{m}  \\
\times  \exp\left[-\frac{h\nu}{kT(\tilde{r}_{m})}-\tau_\nu(\tilde{r}_{m})\right] 
\end{split}
\label{eq:shell}
\end{equation}
where the optical depth $\tau_\nu$ from $r$ to the surface of the cloud is
\begin{equation}
\tau_{\nu}=\tilde{\kappa}_{\nu}\left(\tilde{r}^{-k_{\rho}+1}-
\rct^{-k_{\rho}+1}\right) \; .
\label{eq:tau}
\end{equation}
Equation \ref{eq:shell} (CM05) is the analytic expression that we fit to the observed SEDs, where the quantities $\tilde{\nu}\equiv\nu/\nu_{\rm ch}$, $\tilde{r}_{m}\equiv r_{\rm m}(\nu)/\rch=f(\rct,\tch)$, and 
$\theta_{\rm ch} = \rch/D_{A}$, where $D_{A}$ is the angular diameter distance of the source.
The characteristic frequency $\nu_{\rm ch} \equiv k\tch/h$, and the quantity $\Delta\tilde{r}_{m}$ is a function of the SED variables $\tch$ and $\rct$.  Thus, given three photometric data points we can fit for the unknowns $\tch$, $\rct$ and $\theta_{\rm ch}$ in Equation \ref{eq:shell}.  This analytic form for the SED 
is easily understood as a dust envelope with photospheric radius $\rch$ emitting most of its radiation at $\nu_{\rm ch}$ (close to the peak frequency of the SED).  Each 
frequency channel emanates from radius $r_{m}(\nu)$ with thickness $\Delta \tilde{r}_{m}$; the high frequency photons must also contend with the opacity of the dust envelope 
relative to their point of origin and the Wien cutoff.  CM05 and CM08 have given expressions for the characteristic emission radii of different frequency channels -- Rayleigh-Jeans:  $h\nu/kT << 1$, 
intermediate: $h\nu \sim kT$ and high-frequency: $h\nu/kT \gg 1, \tau > 1$, which comes from the outer parts of the envelope, close to the photosphere, and regions that mediate an effective tug of war between the opacity and temperature gradient of the envelope, respectively.  The characteristic emission radius $r_{\rm m}(\nu)$ was self consistently determined in CM05 by the requirement that the
integrated luminosity equal the total luminosity.

Once the SED variables $\tch$ and $\rct$ are determined, we can solve for the source parameters of the dust envelope, i..e, the light-to-mass ratio ($L/M$) and the surface density ($\Sigma$) using the following equations, which may be inferred from equations 51 and 52 in CM05:
\begin{equation} 
\frac{L}{M}=1.6\left(\frac{3-\krho}{k_{\rho}-1}\right)\kappa_{\nu_0}\left(\frac{\sigma T_{\rm ch}^{\scriptstyle 4+\beta}}{T_0^\beta}\right)\rct^{\scriptstyle k_{\rho}-2.9}
\; ,
\label{eq:LM}
\end{equation}
\begin{equation}
\Sigma=\frac{4(k_{\rho}-1)}{(3-k_{\rho})}\frac{1}{\kappa_{\nu_{0}}}\left({\frac{T_{0}}{T_{\rm
ch}}}\right)^{\beta}\tilde{R_{c}}^{\scriptstyle -(k_{\rho}-1)} \; .
\label{eq:Sigma}
\end{equation} 

\subsection{Inferring the Redshift}

The parameters we infer, $L/M\delta$ and $\Sigma\delta$, depend on redshift
through the dependence of frequency on redshift.  We can express the
redshift dependence of these parameters in a very simple manner.  
Since the luminosity of a dust envelope satisfies $L \propto R^{2} T^{4}$
and the inferred mass is $M \propto \Sigma R^{2}$,
we have: 
\beq
\frac{L}{M\delta} \propto \frac{T^{4}}{\Sigma\delta}  \; 
\eeq
(CM08).  The redshift dependence of the surface density, $\Sigma\delta$, follows
from noting that the optical depth at the observed frequency
must match that at the emitted frequency, since $\tau_\nu$ determines
the shape of the SED, which is invariant: $\kappa(\nu_\obs)(\Sigma\delta)_\obs
=\kappa(\nu_\rest)(\Sigma\delta)_\rest$, so that $\Sigma_\obs/\Sigma_\rest=(\nu_\rest/\nu_\obs)^\beta=
(1+z)^\beta$.
Since $T_{\rm obs}=T_{\rm rest}/(1+z)$, it follows that (CM08):
\beq
\left(\frac{L}{M\delta}\right)_{\rm obs} = \left(\frac{L}{M\delta}\right)_{\rm rest} (1+z)^{-(4+\beta)}\; .
\label{eq:LM_z_easy}
\eeq

We can now use
equation (\ref{eq:LM_z_easy}), setting $\beta=2$ (WD01), to infer the redshift of a dusty
galaxy from its observed value of $L/M\delta$,
which we have shown can be derived from the FIR SED analytically (CM08): 
\beq 1+z_{\rm inf}=\left[\frac{\langle L/M\delta\rangle}{\left(L/M\delta\right)_{\rm obs}}\right]^{1/6}\; ,
\label{eq:zdet}
\eeq 
where $\langle L/M\delta\rangle$ is the typical value in the \her sample; since $L/M\delta$ can range over
an order of magnitude, we use the geometric mean. 
Note that $L$ is the luminosity derived from fitting to the mm-FIR SED.
A simple way to understand why our method yields a more accurate redshift than estimating it from the temperature (or peak of the SED) is as follows.  Our inference of the redshift is proportional to $L/M^{1/6}$.  Alternately, inferring the redshift from fitting a blackbody function to the SED and estimating a temperature from the shift of the peak of the SED is proportional to the temperature itself (i.e., to the template rest wavelength temperature assumed).  It follows from equation \ref{eq:LM} that $L/M \propto T_{\rm ch}^{6} \rct^{-1}$ (for $\krho=2$ and $\beta=2$).  The parameters ($T_{\rm ch}$ and $\rct$) are correlated inversely for a given luminosity (from equation 1), leading to a smaller range in $L/M^{1/6}$ than in the temperature itself, and therefore a more accurate redshift.  This is the case when $\rct$ can be independently determined, which can be done when there is sufficient information on the shape of the SED, as we discuss in \S 4.  

\section{Observations}
\label{sec:observations}

In this study, we used deep PACS 70, 100 and 160$\,\mu$m observations provided by the \textit{Herschel} Space Observatory as part of the PACS Evolutionary Probe (PEP\footnote{http://www.mpe.mpg.de/ir/Research/PEP}) guaranteed time key program (Lutz et al. 2011).
Specifically, we used here the observations of the Great Observatories Origins Deep Survey-North (GOODS-N) and South (GOODS-S) fields, the Lockman Hole (LH) field, the Cosmological evolution survey (COSMOS) field and the lensed fields Abell 2218, Abell 1835, Abell 2219, Abell 2390, MS1054, CL0024 and MS045.
The sample with spectroscopic redshifts is taken from and fully described in Magnelli et al. (2012).

The first aim of this paper is to test the CM FIR
photometric redshift method on a large and more diverse sample than studied by CM08.
To calibrate and test this photometric redshift method we use a sample
of SMGs detected by PACS and with robust spectroscopic estimates from Magnelli et al. (2012).
Here, we restrict our study to the sub-sample of PACS detected
SMGs and we do not use the SPIRE flux densities in the
computation of our photometric redshifts.
Our final sample contains 80 PACS detected SMGs, which have
sub-mm ($850~\micron, 870~\micron$) or (and) mm (1.1 mm) data (Magnelli et al. 2012).
Of these sources, about one-half have robust spectroscopic redshift estimates.
As in CM08, we restrict our SED fitting to 
sources that have at least three photometric data points with $S/N > 3$.

As a second step we apply the CM photometric method to a PACS-detected SMG
sample of the GOODS-S field (i.e. without requiring any spectroscopic
redshift estimate) and derive their SFRD.
This SMG sample corresponds to the AzTEC (1.1 mm) sample of the GOODS-S
field presented in Scott et al. (2010) and which contains 41 sources
detected with S/N$>3.5$.
This AzTEC sample is cross-matched with our PACS catalogue using a
matching radius of 9\arcsec.
We find that 25/41 AzTEC sources are detected in our PACS catalogue.
We note assuming a $S_{850\,\mu{\rm m}}/S_{1100\,\mu{\rm m}}$ ratio of 2.1
(Scott et al. 2008; Austermann et al. 2010), the depth of the AzTEC survey converts
into a 850$\,\mu$m flux of $4.2\,$mJy.
Thus, using our final sample of 25 galaxies, we derive the SFRD of AzTEC and PACS detected sources (i.e.,
down to $\thicksim1.2$~mJy and $\thicksim2.4$~mJy at 100 and 160~$\mu$m
respectively) sources that have $850~\micron$ fluxes typical of "classical" SMGs ($F_{850~\micron} \ga 5~\rm mJy$).  Our results on the SFRD may be uncertain in part due to the PACS non-detections 
of some of the SMGs in the GOODS-S field, as well as the relatively small number of sources.
Therefore, we emphasize that we derive here the SFRD of a sample of infrared bright galaxies
that have extrapolated $850~\micron$ fluxes typical of classical SMGs, specifically the
AzTEC-PACS sources in the GOODS-S field, rather than the SFRD of the SMG population.

\section{Photometric Redshifts From A Sample of $\it{Herschel}$ Sources \label{sec:photoz}}

\begin{figure}[h]
\begin{center}
\includegraphics[scale=0.5]{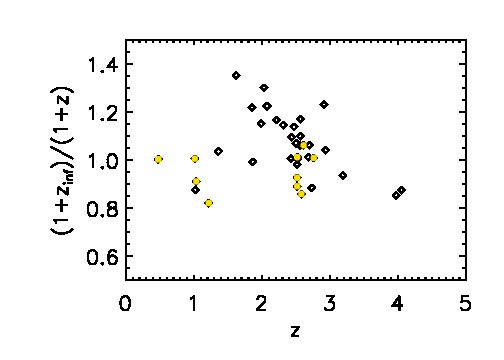}
\caption{Accuracy of redshift determination using the CM photometric redshift method; $z_{\rm inf}$ is the photometric redshift we derive and $z$ is the spectroscopic redshift.  Yellow filled circles mark sources with long wavelength data that sample close to the break frequency.   \label{f:photozacc}}
\end{center}
\end{figure}

We first examine the sub-set with spectroscopic redshifts to verify the accuracy of the CM method.  Some of the sources in this sample are lacking millimeter data, i.e., data points that are close to the break frequency of the SED, where the character of the emission switches from being dominated by hotter photons coming from close to the photosphere to colder photons that come from the outer regions of the envelope. The break frequency of the SED is given by (CM08):
\beq
\nu_{\rm break}=6.35\times 10^{9} \left( \frac{100}{\rct}\right)^{0.4} \frac{\tch}{1+z} \rm Hz \; .
\eeq
For average values of $\tch \sim 100~\rm K$ and $\rct \sim 300$, galaxies at $z \sim 2$ would have a break wavelength of $\sim$ 2 mm.  We have determined (by successively removing long wavelength data points of local ULIRGs and fitting the SEDs) that if the lowest frequency data point is more than a factor of two higher than the break frequency, then the shape of the entire SED (including the transition to intermediate frequencies) can not be clearly determined.  In such cases, we adopt a median value of $\rct$ ($\rct=300$) that is determined by fitting to the sample that has both spectroscopic redshifts and long wavelength millimeter data, and at least three data points with $S/N > 3$ (with the latter being a requirement for all the sources we fit).  We note that the variation in the dust temperature when the SED is fit with a blackbody function is slightly larger than the variation of the characteristic temperature when the SED is fit with our model.  In some cases, as shown in Figure  \ref{f:photozacc}, we do have long wavelength data points that sample the break frequency of the SED.  This allows us to improve the accuracy of our photometric redshift method (as it would allow us to independently fit for not only $\tch$ but also $\rct$), and also provides a better determination of the mass of the envelope.

The average $L/M$ of the sample with spectroscopic redshifts is $30~L_{\odot}/M_{\odot}$, which is comparable to the average $L/M$ determined for SMGs studied in CM08.   This is the template $L/M$ value that we employ in deriving photometric redshifts here. CM08 had noted that the $L/M$ they derived for SMGs was a factor of two lower than local ULIRGs.  The somewhat lower values of $L/M$ for SMGs relative to local ULIRGs found by CM08 is reflected in the lower dust temperatures found in high redshift infrared bright galaxies (Symeonidis et al. 2011).  Figure \ref{f:photozacc} depicts the accuracy of our method for sources in this sample with spectroscopic redshifts.  The yellow filled circles mark the sources that have sufficient long-wavelength data, i.e., $\nu_{\rm min} \la 2\times \nu_{\rm break}$, to determine $\rct$ independently. The average accuracy for the entire sample is 1.11, and the standard deviation is 0.13.  The corresponding numbers when the redshift is inferred by fitting a single temperature blackbody function to the SED are 1.2 and 0.2 respectively.  Here we have used a temperature of 60 K, finding that it yielded the maximal redshift accuracy when fitting with a blackbody function.  Using a modified blackbody fit (with $\beta=1.5$ as recommended by Gordon et al. 2010 and T=50K) yields similar accuracies as for a single temperature blackbody.   The sample with sufficient long-wave data has an average accuracy of 1.07 and standard deviation of 0.075.   The higher accuracy of the CM method is due to fitting for two parameters ($\rct$ and $\tch$) to account for the influence of a range of temperatures on the FIR SED, along with the peak frequency of the SED.  We can fit for both parameters independently when we have data for the long wavelength part of the SED, sampling close to the break frequency.  For the sample that does not have sufficient long-wavelength data, we find a smaller dispersion of characteristic temperatures when we fit our functional form to the data than when we fit a single temperature blackbody, as our functional form is a better approximation to the real FIR SED.  We attribute the slightly higher accuracy of the CM method for the sample with restricted data to the smaller dispersion in temperatures when we fit the data with our functional form.

It is important to contrast our method with other photometric redshift methods that utilize the FIR SED.  We have already discussed above the accuracy in the inference of the redshift by fitting a single temperature blackbody.  Another commonly used FIR redshift method employs the radio-sub-mm spectral index (Carilli \& Yun 1999).  This technique provides a gross indication of the redshift by using the variation of the radio-sub-mm spectral index as a function of redshift.  The uncertainty in redshift for this method depends on the scatter of the spectral index for a given polynomial fit to the data.  The uncertainty in redshift also depends on how well a given polynomial fits the data.  The $\pm 1 \sigma$ variation in the spectral index for the adopted polynomial fit in Carilli \& Yun (1999) is $\sim 1$, with 30\% of the galaxies in their sample falling outside this range. Thus, this method is less accurate than ours.  Aretxaga et al. (2005) sample mock galaxy catalogs with an assumed evolutionary history to derive photometric redshifts from radio to FIR colors, finding a redshift accuracy of $\delta z \sim 0.3$ (rms dispersion about the line $z_{\rm phot}=z_{\rm spec}$).  In terms of our metric $(1+z_{\rm inf})/(1+z_{\rm spec})$, (where we take $z_{\inf}$ to be the most probable photometric redshift) the average accuracy of the sample considered in Aretxaga et al. (2005) is 1.213 (the average accuracy of our sample is 1.12).  Kovacs et al. (2006) suggest the use of the FIR-radio correlation as a photometric redshift estimator, in a manner similar to Carilli \& Yun (1999).  Roseboom et al. (2012) use radio, mm and SPIRE data to find comparable average accuracy (although the standard deviation in their estimates is greater).  In summary, these photometric redshift methods (whether estimating the redshift from Monte Carlo sampling of mock catalogs, the radio-sub-mm spectral index, or from the dust temperature) yield comparable or generally poorer accuracy relative to our method.  

We conclude therefore that our method yields reasonably accurate photometric redshifts for infrared bright galaxies out to $z \sim 5$.  Maximal accuracy of our method can be realized with the addition of long wavelength millimeter data sampling the break frequency, so that $\tch$ and $\rct$ are independently determined.   The addition of near-IR data (Daddi et al. 2009) does improve the accuracy of photometric redshifts beyond what we are able to do here with FIR data alone.  We should note though that our method does not require radio data, as is used in the estimates by Aretxaga et al. (2005), Carilli \& Yun (1999), and Roseboom et al. (2012), or an optical identification, as is used by Daddi et al. (2009).  This may be an advantage compared to other methods (those that use near-IR data) that achieve similar or better accuracy.

\subsection{Cosmic Star Formation History of AzTEC-PACS sources in GOODS-S}

\begin{figure}[h]
\begin{center}
\includegraphics[scale=0.5]{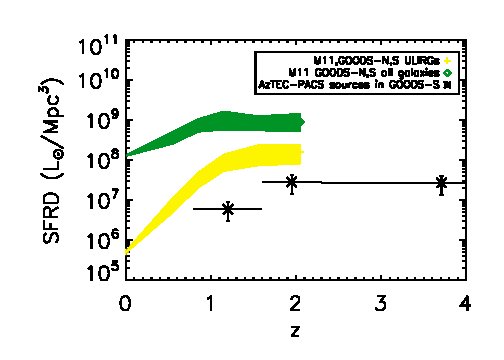}
\caption{Star formation rate density of AzTEC-PACS sources in GOODS-S using \her data (black symbols).  The yellow and green points are the SFRD's derived by Magnelli et al. (2011b) using $24~\micron$ and $70~\micron$ data for ULIRGs and all galaxies respectively in GOODS-S and GOODS-N fields.  The shading indicates the spread in derived values from Magnelli et al. (20011b)   \label{f:SFRD}}
\end{center}
\end{figure}

Now that we have derived the photometric redshifts, we use them to calculate the star formation rate density (SFRD) for the PACS selected SMG sample in the GOODS-S field.  We focus on the GOODS-S field here as that field is homogeneously surveyed down to $5~\rm mJy$ at $850~\micron$ (thereby selecting "classical" SMGs), as discussed \S \ref{sec:observations}.  The other fields (GOODS-N, Lockman Hole, and COSMOS) are
surveyed at different depths over parts of the field in the mm and sub-mm, and as such yield a more heterogeneous sample of galaxies than the GOODS-S field observations.  

We convert our infrared luminosities to star formation rates using the relation presented in Kennicutt (1998), modified (by a factor of $\sim 2$; Portinari et al. 2004) to use the Chabrier IMF (Chabrier 2003), i.e., we take 
\beq
SFR = \frac{L_{\rm IR}}{1 \times 10^{10}L_{\odot}}\; .
\eeq
 The average star formation rate for galaxies with infrared luminosity in excess of $3  \times 10^{12} L_{\odot}$ is $700~M_{\odot}/\rm yr$ for all of the fields we have studied here, including GOODS-S (this is consistent with the observed IRLF at $z \sim 2$, e.g., Wardlow et al. 2010).  This is a factor of $\sim 3$ in excess of the average values found by Dave et al. (2010), and prior to $\it{Herschel}$ observations the total infrared luminosity and hence the star formation rate of SMGs was indeed uncertain.  However, recent $\it{Herschel}$ observations which probe the rest-frame FIR peak of SMGs (Magnelli et al. 2012) do support the higher star formation rates of major merger models (Chakrabarti et al. 2008; Hayward et al. 2011).   
 
Figure 2 shows our derived SFRD (black symbols) over-plotted with the determination by Magnelli et al. (2011b) using 24$~\micron$ and 70$~\micron$ data from the GOODS-S and GOODS-N fields.  The green and yellow symbols depict the SFRD for all galaxies at a given epoch and all ULIRGs respectively.  Note that the sample in Magnelli et al. (2011b) is a more diverse sample than the one we analyze here, including LIRGs and normal galaxies as well as ULIRGs.  We have derived the SFRD shown here (and corresponding infrared luminosity density) using the standard $1/V_{\rm max}$ method (Schmidt 1968).  The luminosity density is given by:
\beq
\phi(\bar{z}_{j}) \Delta z_{j} = \sum_{z_{j-1} \leq z_{i} <z_{j}} \frac{L_{i}}{ \it{V}_{c,i}}\; .
\eeq
Here, $\it{V}_{c,i}=\rm min [ \it{V}_{c}(z_{\rm max,i}) , \it{V}_{c}(z_{j}) ] - \it{V}_{c}(z_{j-1})$, with $\it{V}_{c}$ being the co-moving volume; $i$ labels sources and $j$ redshift bins, so that $z_{\rm max,i}$ is the maximum redshift at which 
source $i$ would have been detected, $\Delta z_{j} = z_{j} - z_{j-1}$, $L_{i}$ is the luminosity of source $i$, and $\bar{z}_{j} = 0.5(z_{j-1}+z_{j})$.    The star formation density in a redshift bin is then simply the luminosity density scaled by the conversion factor given by Kennicutt (1998), using the Chabrier IMF correction.

We find that the sub-mm bright PACs sources contribute 15 \% to the SFRD at $z \sim 2$ relative to ULIRGs (as determined in Magnelli et al. 2011b), and 3 \% relative to the total SFRD at that epoch.  These percentages are relative to the most probable value for the SFRD determined by Magnelli et al. (2011b), with the spread in derived values indicated in Figure \ref{f:SFRD}.  For comparison, Magnelli et al. (2011b) found that the ULIRG population contributes 17 \% to the total SFRD at $z \sim 2$.   As discussed in \S \ref{sec:observations}, the sources we consider here have a $850~\micron$ flux typical of "classical" SMGs ($F_{850~\micron} \ga 5~\rm mJy$).  However, given the relatively small number of sources ($\sim 30$), we interpret our findings as yielding the SFRD of AzTEC-PACS sources in GOODS-S, rather than that of the SMG population.  It is worthwhile noting that there are systematic differences in the determination of the SFRD of SMGs using different surveys.  For instance, radio-detected LESS SMGs in the Extended Chandra Deep Field South (ECDFS) as studied by Wardlow et al. (2010) have a different redshift evolution and overall contribution to the SFRD than the sample studied by Chapman et al. (2005).  Our results are comparable with the SFRD of mm-selected sources in the Lockman Hole and GOODS-N fields calculated by Roseboom et al. (2012).  Our determination of the SFRD of the AzTEC-PACS sources is a lower limit to the ULIRG population which are a less extreme version of infrared bright galaxies and increasingly common at high redshifts (Wuyts et al. 2011).  The shape of the SFRD of the AzTEC-PACS sources does not show a decline out to $z \sim 4$.  

\section{Discussion}

There are a few caveats worth mentioning, relating to the use of our method and astrophysical effects.  If lensed galaxies are present in our sample, they would artificially boost the number counts of SMGs.  Negrello et al. (2007) estimate the probability of lensing as a function of the sub-mm flux of a galaxy, which we can use to ascertain the likelihood of galaxies in our sample being lensed.  They find a significant incidence of lensing ($\ga 50 \%$) for galaxies with $850~\micron$ fluxes in excess of 40$~\rm mJy$, but all our sources are well below this threshold.   For \textit{known} lensed galaxies, we have checked that our method works equally well on the lower luminosity SMGs in the A2218 lensed cluster.  We conclude therefore that lensing is not a serious source of concern for this sample, i.e., the probability of galaxies being lensed in the fields we have studied is very low given their $850~\micron$ flux.  Moreover, we have checked that for known lensed galaxies in the A2218 lensing cluster, our method yields robust photometric redshifts.

Our simple model is designed to yield through SED-fitting the large-scale parameters of infrared bright galaxies, namely the mass, size, and luminosity once the redshift is determined using the CM photometric redshift method.  We do not treat the effects on the SED of a clumpy dust envelope or distributed sources of luminosity.   Although we do not treat the morphology of infrared bright galaxies in detail, our derivation of the mass, size, luminosity and density profile is in good agreement with observational determinations (CM08).  The advantage of our model is that it incorporates a self-consistent range of dust temperatures in an analytic expression for the emitted luminosity, which we have shown can be applied to infrared bright galaxies to robustly derive redshifts and large-scale source parameters out to $z \sim 4$.

\section{Conclusion}

$\bullet$ We have applied the photometric redshift method of CM08 to \her data to demonstrate the accuracy of our method for a large and diverse sample of SMGs.  We find accuracies of $\sim 10 \%$ relative to spectroscopic redshifts, i.e., in $(1+z_{\rm inf})/(1+z)$, out to $z \sim 4$.   

$\bullet$ We also give the average values for the $L/M$ ratios of SMGs in these samples, which is lower than that for local ULIRGs by a factor of 2.  Our average star formation rate is $700~M_{\odot}/\rm yr$ for galaxies with luminosities in excess of $3 \times 10^{12} L_{\odot}$.

$\bullet$ We estimate the star formation rate density of sub-mm bright PACS sources in the GOODS-S field.  These sources have an extrapolated $850~\micron$ flux typical of classical SMGs,  $F_{850~\micron} \ga 5 ~\rm mJy$ (we extrapolated the sub-mm flux from AzTEC observations as discussed in \S \ref{sec:observations}).   Our derivation of the SFRD is a lower limit, particularly at low redshifts ($z < 1$) where normal spirals (and secular processes or minor tidal interactions) drive the star formation history of the universe.   These PACS sources contribute 15 \% to the SFRD relative to ULIRGs at $z \sim 2$, and 3 \% relative to the total SFRD at that epoch that is produced by all galaxies.  We find that there is no decline in the shape of the SFRD of the sub-mm bright PACS sources in the GOODS-S field out to $z \sim 4$.

\bigskip
\bigskip

\acknowledgements
SC thanks  Max-Planck-Institut f\"ur Extraterrestrische Physik for their hospitality during the initial study of the $\it{Herschel}$ sample.  We also thank Reinhard Genzel for helpful discussions.  PACS has been developed by a consortium of institutes led by MPE
(Germany) and including UVIE (Austria); KU Leuven, CSL, IMEC (Belgium);
CEA, LAM (France); MPIA (Germany); INAF-IFSI/OAA/OAP/OAT, LENS,
SISSA (Italy); IAC (Spain). This development has been supported by the funding
agencies BMVIT (Austria), ESA-PRODEX (Belgium), CEA/CNES (France),
DLR (Germany), ASI/INAF (Italy), and CICYT/MCYT (Spain).  The research of CFM is supported in part by the NSF through grant AST 0908553.


\references

Austermann, J.~E., Dunlop, J.~S., Perera, T.~A., {et~al.} 2010, \mnras, 401, 160 

Blain, A., Smail, I., et al., 2002, PhR, 369, 111B 

Chabrier, G., 2003, PASP, 115, 763 

Chakrabarti, S., \& McKee, C.F.M., 2005, ApJ, 631, 792C [CM05] 

Chakrabarti, S. \& McKee, C.F.M., 2008, ApJ, 683, 693C [CM08]  

Chakrabarti, S., Fenner, Y., et al., 2008, ApJ, 688, 972C  

Chakrabarti, S., \& Whitney, B.A., 2009, ApJ, 690, 1432C 

Chapman, S.C., Smail, I., et al., 2004, ApJ, 614, 671C 

Chapman, S.C., Blain, A.W., et al., 2005, ApJ, 622, 772 

Dave, R.., Finlator, K., et al., 2010, MNRAS, 404, 1355D 

Hayward,C., Narayanan, D., et al. 2011, ASPC, 440, 369H 

Hayward,C., Keres, D., et al. 2011, ApJ, 743, 159H 

Hopkins, A., 2004, ApJ, 615, 209 

Kennicutt, R.C., 1998, ApJ, 498, 541 

Kovacs, A., Chapman, S.C., et al. 2006, ApJ, 650, 592K 

Lacy, M., Storrie-Lombardi, L. J., et al., 2004, ApJS, 154, 166L 

Madau, P., Pozzetti, L. \& Dickinson, M., 1998, ApJ, 498, 106 

Magnelli, B., Lutz, D., et al., 2010, A\&A, 518L, 28M 

Magnelli, B., Elbaz D., et al., 2011b, A\&A, 528A, 35M  

Magnelli, B., Lutz, D., et al., 2012, accepted to A\&A,arXiv:1202.0761 

Michalowski, M., Hjorth, J., \& Watson, D., 2010, A\&A, 514A, 67M 

Pilbratt, G.L., Riedinger, J.R., et al., 2010, A\&A, 518L, 1P 

Poglitsch,, A., Waelkens, C., et al., 2010, A\&A, 518L, 2P 

Sanders, D. B., Soifer, B.T., et al., 1988, ApJ, 325, 74S 

Schmidt, M., 1968, ApJ, 151, 393 

Springel, V., Di Matteo, T., \& Hernquist, L., 2005, MNRAS, 364, 1105S 

Toomre, A., \& Toomre, J., 1972, ApJ, 178, 623 

Wardlow, J., L., et al. 2010, MNRAS, 201, 2299 

Weingartner, J., \& Draine, B., 2001, ApJ, 548, 296 [WD01] 

Wuyts, S., et al., 2011, ApJ, 738, 106 

\end{document}